\documentclass[amsmath,amssymb,eqsecnum,notitlepage,nofootinbib,a4paper]{revtex4-1}
\usepackage{graphicx}
\usepackage{fontenc}
\usepackage{subfigure}
\usepackage{setspace}
\usepackage[breaklinks, colorlinks, citecolor=magenta!90!black, linkcolor=blue!90!black ]{hyperref}
\usepackage{amsmath}
\usepackage{amssymb}
\usepackage{bm}
\usepackage{verbatim}
\usepackage{xcolor}

\graphicspath{{./}}

\def\half{\frac{1}{2}}
\def\third{\frac{1}{3}}
\def\quarter{\frac{1}{4}}
\def\0{\varnothing} 
\def\T{\mathsf{T}} 
\def\BI{\gamma} 
\def\R{\mathcal{R}}
\def\K{\mathcal{K}}
\def\H{\mathcal{H}}

\def\xty{\left({x}\leftrightarrow{y}\right)} 

\newcommand{\sgn}[1]{\operatorname{sgn} (#1)}

\newcommand{\partdif}[2]{\frac{\partial {#1}}{\partial {#2}}}
\newcommand{\funcdif}[2]{\frac{\delta {#1}}{\delta {#2}}}
\newcommand{\totdif}[2]{\frac{\mathrm{d} {#1}}{\mathrm{d} {#2}}}
\newcommand{\figref}[1]{Fig.~\ref{#1}}

\newcommand{\secref}[1]{section~\ref{#1}}
\newcommand{\appref}[1]{appendix~\ref{#1}}

\begin{document}

\begin{flushleft}
KCL-PH-TH/2019-08
\end{flushleft}

\title{The general gravitational Lagrangian with deformed covariance}

\author{Rhiannon Cuttell} \email{rhiannon.cuttell@kcl.ac.uk}
\author{Mairi Sakellariadou} \email{mairi.sakellariadou@kcl.ac.uk}
\affiliation{
    Department of Physics,
	King's College London,
	University of London,
	Strand,
	London,
	WC2R 2LS,
	U.K.
}
\date{\today}

\begin{abstract}
We derive the gravitational Lagrangian to all orders of curvature when the canonical constraint algebra is deformed by a phase space function as predicted by some studies into loop quantum cosmology.
The deformation function seems to be required to satisfy a non-linear equation usually found in fluid mechanics, and can form discontinuities quite generally.
\end{abstract}

\maketitle


\tableofcontents 


\section{Introduction}
\label{sec:intro}

Many effective models of quantum gravity work from the hypothesis that the symmetries of general relativity should be deformed by quantum effects.  The mechanism by which it is implemented at the effective level is diverse and often only considers individual particles.
This includes deformed special relativity \cite{AmelinoCamelia:2000mn} and rainbow gravity \cite{Magueijo:2002xx}, which struggle to go beyond describing particles coupled to a metric dependent on the particle's energy.  Such models can suffer from a breakdown of causality \cite{hossenfelder_box-problem_2009}, or find it difficult to describe multi-particle states \cite{hossenfelder_multi-particle_2007}.  The model we consider in this paper, deformed general relativity, should not suffer from these problems by construction since it is energy density and curvature which the deformation depends on.

A specific kind of deformation consistently appears in some investigations of loop quantum cosmology, when loop quantisation effects are introduced into mini-superspace models without causing anomalies \cite{Bojowald:2008gz, Bojowald:2008bt, Perez:2010pm, Mielczarek:2011ph, Cailleteau2012a, Mielczarek:2012pf, Cailleteau2013}.  
The constraint algebra\footnote{also known as the hypersurface deformation algebra}, which ensures space-time covariance is maintained when we have made a space-time decomposition \cite{teitelboim_how_1973}, is deformed by a phase space function $\beta(q,p)$.
For a more in-depth review, please see ref.~\cite{cuttellthesis}.

In this paper seek to derive the most general effective gravitational action which satisfies the deformed constraint algebra.  We derive the restrictions on the Lagrangian in \secref{sec:dist-eqn_sol}.  In \secref{sec:allact} we use them to find the allowed forms of the deformation and the general Lagrangian.  Curiously, we find the deformation function must satisfy the inviscid form of the Burgers' equation in curvature space.  This may be related to the curved phase space hypothesis \cite{AmelinoCamelia:2011bm}, which is known to be linked with similar models of deformed relativity.

These calculations generalise those presented in ref.~\cite{cuttell2014} where the fourth order gravitational Lagrangian was perturbatively derived from the constraint algebra%
\footnote{an updated version of the calculation of the fourth order perturbative gravitational Lagrangian can be found in \cite{cuttellthesis}\label{ft:updatenote}}.
This is a companion paper to ref.~\cite{cuttellconstraint}, wherein we calculate the general scalar-tensor Hamiltonian with deformed general covariance.


\subsection{Space-time decomposition}
\label{sec:intro_decomposition}

We foliate the bulk space-time manifold $\mathcal{M}$ into a stack of labelled spatial hypersurfaces, $\Sigma_t$ \cite{bojowald2010canonical, Arnowitt:1962hi, Gourgoulhon:2007ue}.
The time-evolution is described by the time vector $t^a$.
Each spatial slice has a metric $q_{ab}$, and a normal vector $n^a$, so we can project the time vector into its normal and tangential components.  This produces the lapse ${N=-n_{a}t^{a}}$ and shift ${N^{a}=q^{a}_{b}t^{b}}$.  These act as Lagrange multipliers in the classical action, and they produce constraints (i.e. they vanish in the dynamical regime) definable from the total Hamiltonian,
\begin{equation}
    C : = \funcdif{ H }{ N },
\quad
    D_a : = \funcdif{ H }{ N^a },
\end{equation}
which are respectively known as the Hamiltonian constraint, and the diffeomorphism constraint.
The classical Poisson bracket structure of these constraints forms a Lie algebroid \cite{Bojowald:2016hgh},
\begin{subequations}
\begin{align}
    \big\{ D_a [N^a], D_b [M^b] \big\} & = D_a \big[ \mathcal{L}_M N^a \big],
        \label{eq:con-alg_DD}
        \\
    \big\{ C [N], D [M^a] \big\} & = C \big[ \mathcal{L}_M N \big],
        \label{eq:con-alg_CD}
        \\
    \big\{ C [N], C [M] \big\} & = D_a \big[ \, q^{ab} \left( N \partial_b M - \partial_b N M \right) \big].
        \label{eq:con-alg_CC}
\end{align}
    \label{eq:con-alg}%
\end{subequations}
Since there are no anomalous terms (those not constrained to vanish), $N$ and $N^a$ are gauge functions which do not affect the observables, and therefore the spatial slicing does not affect the dynamics.  
As interpreted in ref.~\cite{teitelboim_how_1973}, \eqref{eq:con-alg_DD} shows that $D_a$ is the generator of spatial morphisms, \eqref{eq:con-alg_CD} shows that $C$ is a scalar density of weight one, and \eqref{eq:con-alg_CC} specifies the form of $C$ such that it ensures the embeddability of the spatial slices in space-time geometry.



Classical general relativity with a space-time decomposition can be formulated equivalently using different variables.  Geometrodynamics usually uses the spatial metric $q_{ab}$ and the extrinsic curvature $K_{ab}=\half\mathcal{L}_nq_{ab}$.
Connection dynamics uses the Ashtekar-Barbero connection $A^I_a$ and densitised triads $E^a_I$ \cite{Ashtekar:1986yd, Barbero:1994ap}. 
The connection has an ambiguity in its definition given by the Barbero-Immirzi parameter $\gamma$, which parameterises the contribution of the extrinsic curvature relative to the triad's spin connection, but the exact value of $\BI$ should not affect the dynamics \cite{Immirzi:1996dr}.
The other prominent alternative is loop dynamics, which uses integrated versions of the $A^I_a$ and $E^a_I$.  If the integration regions are taken to be infinitesimal, then the one can easily relate loop dynamics and connection dynamics \cite[p.~21]{Rovelli:2014ssa}.

When each set of variables is quantised, they are no longer equivalent, and $\BI$ does now affect the dynamics \cite{Ashtekar:1997yu, Brahma:2016tsq}. For complex $\BI$, care has to be taken to make sure the classical limit is real general relativity, rather than complex general relativity.  Significantly, quantising loop variables (loop quantum gravity) explicitly discretises the geometry, and so the integration regions cannot be taken to be infinitesimal \cite[p.~105]{Rovelli:2014ssa}.
In this work, we choose to use metric variables of geometrodynamics to build a semi-classical model of gravity. This is because the comparison to classical gravity models should be clearer, and there is no ambiguity arising from $\BI$.

We are considering only the spatial metric field $q_{ab}$ and its normal derivative $v_{ab}=\mathcal{L}_nq_{ab}$.
Time derivatives above first-order, mixed-type derivatives such as $\nabla_c{v}_{ab}$, and tensor contractions of derivatives above second order are associated with additional degrees of freedom \cite{Deruelle2010}, and for simplicity we do not consider such terms in this paper.
The only covariant quantities we can form up to second order in derivatives from the spatial metric are the determinant $q=\det{q_{ab}}$ and the Ricci curvature scalar $R$.  The normal derivative can be split into its trace and traceless components, ${v_{ab}=v^\T_{ab}+\third{}vq_{ab}}$, so it can form scalars from the trace $v$ and a variety of contractions of the traceless tensor $v^\T_{ab}$.  However, to second order we only need to consider ${w:=v^\T_{ab}v_\T^{ab}}$.
Therefore, we consider the Lagrangian given by ${L=L(q,v,w,R)}$.


\subsection{Deformed constraint algebra}
\label{sec:intro_algebra}

Some models of loop quantum cosmology predict that the symmetries of general relativity should be deformed in a specific way in the semi-classical limit \cite{Bojowald:2008gz, Bojowald:2008bt, Perez:2010pm, Mielczarek:2011ph, Cailleteau2012a, Mielczarek:2012pf, Cailleteau2013}.  This appears from incorporating loop variables in a mini-superspace model while keeping the constraint algebra anomaly-free but allowing counter-terms to deform the classical form of the algebra.
This retains the gauge invariance of the theory and the arbitrariness of the lapse and shift.  
If anomalous terms \emph{were} to appear in the constraint algebra, then the gauge invariance would be broken and the constraints could only be solved at all times for specific $N$ or $N^a$.  This means that there would a privileged frame of reference, and therefore no general covariance.

In the referenced studies, it is strongly indicated that the bracket of two Hamiltonian constraints \eqref{eq:con-alg_CC} is deformed by a phase space function $\beta$,
\begin{equation}
    \{ C [N] , C [M] \} = D_a [ \beta q^{ab} \left( N \partial_b M - \partial_b N M \right) ].
    \label{eq:con-alg_def}
\end{equation}
This has not been shown generally, but has been shown for several models independently.
There are no anomalies in the constraint algebra, so a form of general covariance is preserved.  However, it may be that the interpretation of a spatial manifold evolving with time being equivalent to a foliation of space-time (also known as `embeddability') is no longer valid.

These deformations only appear to be necessary for models when the Barbero-Immirzi parameter $\BI$ is real.  For self-dual models, when $\BI=\pm{i}$, this deformation does not appear necessary \cite{BenAchour:2016brs}.  However, self-dual variables are not desirable in other ways.  They do not seem to resolve curvature singularities as hoped, and obtaining the correct classical limit is non-trivial \cite{Brahma:2016tsq}.  Because of this, even though we use metric variables in this study, considering $\beta\neq1$ and ensuring the correct classical limit means there should be relevance to the models of loop quantum cosmology with real $\BI$.



From the constraint algebra, we are able to find the specific form of the Hamiltonian constraint $C$ or the Lagrangian $L$ for a given deformation $\beta$.  The diffeomorphism constraint $D_a$ is not affected when the deformation is a scalar%
\footnote{That is, when $\beta$ has a density weight of zero \cite{bojowald2010canonical}.}
and so is completely determined as shown in ref.~\cite{cuttellconstraint}.  With $D_a$ and $\beta$ as inputs, we can find $C$ and thereby $L$ by manipulating \eqref{eq:con-alg_def}.

Firstly, we must find the unsmeared form of the deformed algebra.  At this point we do not need to specify the variables, and leave them merely as ${(q_I,p_I)}$
\begin{subequations}
\begin{align}
    0 & = \big\{ C [N] , C [M] \big\} - D_a \big[ \beta q^{ab} \left( N \partial_b M - \partial_b N M \right) \big],
\\ & 
    = \int \mathrm{d}^3 z \left\{ 
        \sum_I \funcdif{ C [N] }{ q_I(z) } \funcdif{ C [M] }{ p_I (z) }
        - \left( D^a \beta N \partial_a M \right)_z
    \right\}
    - \left( N \leftrightarrow M \right).
    \label{eq:con-alg_def_expanded}
\end{align}
\end{subequations}
For when we wish to derive the action instead of the constraint, we can transform the equation by first noting that,
\begin{equation}
    \funcdif{ C [N] }{ q_I } = - \funcdif{ L [N] }{ q_I },
\quad
    N v_I = \funcdif{ C [N] }{ p_I },
\end{equation}
where $v_I:=\mathcal{L}_n{q_I}$, and the Lagrangian is defined such that
$\textstyle{S = \int \mathrm{d} t \mathrm{d}^3 x N L = \int \mathrm{d} t L [N]}$.
We substitute these into \eqref{eq:con-alg_def_expanded}, then take functional derivatives to remove $N$ and $M$,
\begin{equation}
    0 = \sum_I \funcdif{ L (x) }{ q_I (y) } v_I (y)
    + \left( \beta D^a \partial_a \right)_x \delta \left( x, y \right)
    - \xty.
    \label{eq:dist-eqn_act}
\end{equation}
To find a useful form for this, we need to use a specific form for the diffeomorphism constraint.  Because it depends on momenta, we must replace them using,
\begin{equation}
    p_I := \funcdif{ S }{ \dot{q_I} } = \frac{1}{N} \funcdif{ L [N] }{ v_I },
    \label{eq:momentum_def_full}
\end{equation}
and, as before, if we note that we will only consider actions without mixed derivatives this simplifies to
\begin{equation}
    p_I = \partdif{ L }{ v_I }.
    \label{eq:momentum_def}
\end{equation}
Therefore, substituting the diffeomorphism constraint, and the momenta \eqref{eq:momentum_def} into \eqref{eq:dist-eqn_act}, we find the distribution equation which can be used for restricting the form of the deformed action.



We can determine the relationship between the order of the deformation function and the order of the associated action by comparing orders of time derivatives.
Consider the distribution equation \eqref{eq:dist-eqn_act} with only a scalar field,
\begin{equation}
\begin{split}
    0 & =
    \funcdif{ L (x) }{ \psi (y) } v_\psi (y)
    + \left( \beta \partdif{ L }{ v_\psi } \partial^a \psi \partial_a \right)_x \delta ( x, y )
    - \xty,
\end{split}
\end{equation}
where we have used the diffeomorphism constraint for a scalar field ${D_a=p_\psi\partial_a\psi}$ \cite{cuttellconstraint}, where $\textstyle{p_\psi=\partdif{L}{v_\psi}}$.
Let us consider a simplified model to match the derivative orders for the deformation and the derivative orders for the Lagrangian in a way analogous to dimensional analysis.  First order time derivatives are given by $v_{\psi}$ and two orders of spatial derivatives are given by $\Delta$.  We can collect terms in the distribution equation of the same order of time derivatives as they are linearly independent.  Schematically, the distribution equation is given by,
\begin{equation}
    0 = \partdif{L}{\Delta} v_\psi + \partdif{L}{v_\psi} \beta,
\end{equation}
and expanding the Lagrangian and deformation in powers of $v_\psi$, 
\begin{equation}
    L = \sum_{m=0}^{n_L} L^{(m)} v_{\psi}^m,
\quad
    \beta = \sum_{m=0}^{n_\beta} \beta^{(m)} v_{\psi}^m,
\end{equation}
the coefficient of $v_{\psi}^n$ is then given by,
\begin{equation}
    0 = \partdif{ L^{(n-1)} }{ \Delta }
    + \sum_{m=0}^{n_\beta} \left( n - m + 1 \right) L^{(n-m+1)} \beta^{(m)}.
\end{equation}
We can relabel and rearrange to find a schematic solution for the highest order of $L$ appearing here,
\begin{equation}
    L^{(n)} =
    \frac{ -1 }{ n \beta^{ (0) } }
    \left\{
        \partdif{ L^{ (n-2) } }{ \Delta }
        + \sum_{m=1}^{n_\beta} \left( n - m \right) \beta^{(m)} L^{(n-m)}
    \right\}.
\end{equation}
We can see that if $n_\beta>0$, then this equation is recursive and $n_L\to\infty$ because there is no natural cut-off, suggesting that a deformed $L$ is required to be non-polynomial.  If we wish to truncate the action at some order, then it must be treated as an perturbative approximation.  We considered a perturbative fourth order Lagrangian in ref.~\cite{cuttell2014}${^{\ref{ft:updatenote}}}$, and the non-perturbative gravitational Lagrangian is considered in this paper.


\section{Solving the distribution equation}
\label{sec:dist-eqn_sol}

The general deformed Lagrangian must satisfy the distribution equation from \eqref{eq:dist-eqn_act}, which when we are only considering metric variables is given by
\begin{equation}
    0 = 
    \funcdif{ L (x) }{ q_{ab} (y) } v_{ab} (y)
    + \left( \beta D^a \partial_a \right)_x \delta \left( x, y \right)
    - \xty.
    \label{eq:dist-eqn}
\end{equation}
As shown in ref.~\cite{cuttellconstraint} the diffeomorphism constraint for a metric is uniquely given by,
\begin{equation}
    D^a = - 2 \nabla_b p^{ab}
    = - 2 \left( \delta^a_{(b} \partial_{c)} + \Gamma^a_{bc} \right) \partdif{ L }{ v_{bc} }.
    \label{eq:diffeomorphism}
\end{equation}
Firstly, we integrate \eqref{eq:dist-eqn} by parts to move spatial derivatives from $L$ onto the delta functions.  We discard the surface term and find,
\begin{equation}
    0 = \funcdif{ L (x) }{ q_{ab} (y) } v_{ab} (y)
    - 2 \left( \beta \partdif{ L }{ v_{bc} } \Gamma^a_{bc} \partial_a \right)_x \delta ( x, y )
    + 2 \left( \partdif{ L }{ v_{ab} } \partial_b \right)_x \big[ \left( \beta \partial_a \right)_x \delta ( x, y ) \big]
    - \xty,
\end{equation}
from this we take the functional derivative with respect to $v_{ab}(z)$ (after relabelling the other indices),
\begin{equation}
\begin{gathered}
    0  =
    \funcdif{ L (x) }{ q_{ab} (y) } \delta ( y, z )
    + \left\{
        \funcdif{ \partial L (x) }{ q_{cd} (y) \partial v_{ab} (x) } v_{cd} (y)
        + 2 \left( \partdif{ \beta_{,d} }{ v_{ab,e} } \partdif{ L }{ v_{cd} } \partial_{c} \right)_x \delta ( x, y ) \partial_{d(x)}
\right. \\ \left.
        + 2 \left[ 
            \partdif{}{v_{ab}} \left( \partial_d \beta \partdif{ L }{ v_{cd} } - \beta \partdif{ L }{ v_{de} } \Gamma^c_{de} \right) \partial_c 
            + \partdif{}{ v_{ab} } \left( \beta \partdif{ L }{ v_{cd} } \right) \partial_{cd}
        \right]_x \delta ( x, y )
    \right\} \delta ( x, z )
    - \xty.
\end{gathered}
\end{equation}
We move the derivative from $\delta(x,z)$ and exchange some terms using the $\xty$ symmetry to find it in the form,
\begin{equation}
    0 = A^{ab} ( x , y ) \delta( y , z ) - A^{ab} ( y , x ) \delta( x , z ),
    \label{eq:dist-eqn_Aab}
\end{equation}
where,
\begin{equation}
\begin{split}
    A^{ab} ( x , y ) & = \funcdif{ L (x) }{ q_{ab} (y) }
    - v_{cd} (x) \funcdif{ \partial L (y) }{ q_{cd} (x) \partial v_{ab} (y) }
    + 2 \left\{ \partdif{}{ v_{ab} } 
        \left( \beta \partdif{ L }{ v_{de} } \Gamma^c_{de} - \partial_d \beta \partdif{ L }{ v_{cd} } \right) \partial_c
\right. \\ & \left.
    - \partdif{}{ v_{ab} } \left( \beta \partdif{ L }{ v_{cd} } \right) \partial_{cd}
        + \partial_e \left( \partdif{ \beta_{,d} }{ v_{ab,e} } \partdif{ L }{ v_{cd} } \right) \partial_c
    \right\}_y \delta ( y, x ).
\end{split}
\end{equation}
Integrating over $y$, we find that part of the equation can be combined into a tensor dependent only on $x$,
\begin{equation}
\begin{split}
    0 & = A^{ab}(x,z) - \delta(z,x) \int \mathrm{d}^3 y A^{ab}(y,x),
        \\
    & = A^{ab}(x,z) - \delta(z,x) A^{ab} (x),
    \quad \mathrm{where} \;
    A^{ab}(x) = \int \mathrm{d}^3 y A^{ab} \left(y, x \right).
\end{split}%
\end{equation}
Substituting in the definition of $A^{ab}(x,z)$ then relabelling,
\begin{equation}
\begin{split}    
    0 &= \funcdif{ L (x) }{ q_{ab} (y) }
    - v_{cd} (x) \funcdif{ \partial L (y) }{ q_{cd} (x) \partial v_{ab} (y) }
    + 2 \left\{ \partdif{}{ v_{ab} } 
        \left( \beta \partdif{ L }{ v_{de} } \Gamma^c_{de} - \partial_d \beta \partdif{ L }{ v_{cd} } \right) \partial_c
\right. \\ & \left.
    - \partdif{}{ v_{ab} } \left( \beta \partdif{ L }{ v_{cd} } \right) \partial_{cd}
        + \partial_e \left( \partdif{ \beta_{,d} }{ v_{ab,e} } \partdif{ L }{ v_{cd} } \right) \partial_c
    \right\}_y \delta ( y, x )
    - A^{ab} (x) \delta ( x, y ).
\end{split}
\end{equation}
To find this in terms of one independent variable, we multiply by the test tensor $\theta_{ab}(y)$, and integrate by parts over $y$. Then collecting derivatives of $\theta_{ab}$,
\begin{equation}
\begin{split}
    0 & = \theta_{ab} \left( \cdots \right)^{ab}
    + \partial_c \theta_{ab} \left\{
        \partdif{ L }{ q_{ab,c} } 
        + v_{de} \partdif{^2 L }{ q_{de,c} \partial v_{ab} } - 2 v_{ef} \partial_d \left( \partdif{^2 L }{ q_{ef,cd} \partial v_{ab} } \right)
\right. \\ & \left.
        + 2 \partdif{}{ v_{ab} } \left( \partial_d \beta \partdif{ L }{ v_{cd} } - \beta \partdif{ L }{ v_{de} } \Gamma^c_{de} \right)
        - 4 \partial_d \left[ \partdif{}{ v_{ab} } \left( \beta \partdif{ L }{ v_{cd} } \right) \right]
        + 2 \partial_e \left( \partdif{ \beta_{,d} }{ v_{ab,c} } \partdif{ L }{ v_{de} } \right)
    \right\}
\\ & 
    + \partial_{cd} \theta_{ab} \left\{ 
        \partdif{ L }{ q_{ab,cd} }
        - v_{ef} \partdif{^2 L }{ q_{ef,cd} \partial v_{ab} }
        - 2 \partdif{}{ v_{ab} } \left( \beta \partdif{ L }{ v_{cd} } \right)
        + 2 \partdif{ \beta_{,e} }{ v_{ab,(c} } \partdif{ L }{ v_{d)e} }
    \right\},
\end{split}
\end{equation}
where we have discarded the terms containing $\theta_{ab}$ without derivatives, because they do not provide any restrictions on the form of the Lagrangian.
This is simplified by noting that $\partial_c$ and ${\partdif{}{v_{ab}}}$ commute, and that 
${\partdif{\beta_{,e}}{v_{ab,c}}=\delta^c_e\partdif{\beta}{v_{ab}}}$.  
Therefore, the solution is given by,
\begin{equation}
\begin{split}
    0 & = \theta_{ab} \left( \cdots \right)^{ab}
    + \partial_c \theta_{ab} \left\{
        \partdif{ L }{ q_{ab,c} } 
        + v_{de} \partdif{^2 L }{ q_{de,c} \partial v_{ab} } - 2 v_{ef} \partial_d \left( \partdif{^2 L }{ q_{ef,cd} \partial v_{ab} } \right)
        - 2 \Gamma^c_{de} \partdif{}{ v_{ab} } \left( \beta \partdif{ L }{ v_{de} } \right)
\right. \\ & \left.
        - 2 \partial_d \beta \partdif{^2 L }{ v_{ab} \partial v_{cd} }
        - 4 \beta \partial_d \left( \partdif{^2 L }{ v_{ab} \partial v_{cd} } \right)
        - 2 \partdif{ \beta }{ v_{ab} } \partial_d \left( \partdif{ L }{ v_{cd} } \right)
    \right\}
\\ &
    + \partial_{cd} \theta_{ab} \left\{ 
        \partdif{ L }{ q_{ab,cd} }
        - v_{ef} \partdif{^2 L }{ q_{ef,cd} \partial v_{ab} }
        - 2 \beta \partdif{^2 L }{ v_{ab} \partial v_{cd} }
    \right\}.
\end{split}
    \label{eq:dist-eqn-sol}
\end{equation}
To find the derivatives with respect to spatial derivatives of the metric, we must use equations from ref.~\cite{cuttellconstraint} for decomposing the Ricci curvature scalar.


\subsection{Finding the conditions on the Lagrangian}
\label{sec:conditions}

Substituting the variables into \eqref{eq:dist-eqn-sol}, the resulting equation contains a series of unique tensor combinations.  The test tensor $\theta_{ab}$ is completely arbitrary so the coefficient of each unique tensor contraction with it must independently vanish.
For example, suppose that ${0=B^{ab}\theta_{ab}}$. If $B^{ab}$ can be decomposed in terms of $q_{ab}$ and $v^\T_{ab}$, we find,
\begin{equation}
    0 = q^{ab} \theta_{ab} B_0 + v_\T^{ab} \theta_{ab} B_1 + v_\T^{ac} v_\T^{bd} q_{cd} \theta_{ab} B_2 + v_\T^{ac} v_\T^{bd} v^\T_{cd} \theta_{ab} B_3 + \ldots
\end{equation}
For this to be satisfied for general metrics, each coefficient $B_I$ must vanish independently.

Firstly, we focus on the terms depending on the second order derivative $\partial_{cd}\theta_{ab}$. We evaluate each individual term in \appref{sec:extras}.
Substituting \eqref{eq:d2theta_components} into \eqref{eq:dist-eqn-sol}, we find the following independent conditions,
\begin{subequations}
\begin{align}
    q^{ab} \partial^2 \theta_{ab} : 0 & =
    \partdif{ L }{ R } 
    - \frac{2v}{3} \partdif{^2 L }{ R \partial v }
    + 2 \beta \left( \partdif{^2 L }{ v^2 } - \frac{2}{3} \partdif{ L }{ w } \right),
\label{eq:d2theta_1} \\
    q^{ac} q^{bd} \partial_{cd} \theta_{ab} : 0 & =
    \partdif{ L }{ R } - 4 \beta \partdif{ L }{ w },
\label{eq:d2theta_2} \\
    q^{ab} v_\T^{cd} \partial_{cd} \theta_{ab} : 0 & =
    \partdif{^2 L }{ R \partial v }
    + 4 \beta \partdif{^2 L }{ w \partial v },
\label{eq:d2theta_3} \\
    v_\T^{ab} \partial^2 \theta_{ab} : 0 & =
    \frac{v}{3} \partdif{^2 L }{ R \partial w } 
    - \beta \partdif{^2 L }{ v \partial w },
\label{eq:d2theta_4} \\
    v_\T^{ab} v_\T^{cd} \partial_{cd} \theta_{ab} : 0 & =
    \partdif{^2 L }{ R \partial w }
    + 4 \beta \partdif{^2 L }{ w^2 }.
\label{eq:d2theta_5}
\end{align}
    \label{eq:d2theta}%
\end{subequations}
Before we analyse these equations, we will find the conditions from the first order derivative part of \eqref{eq:dist-eqn-sol}.  There are many complicated tensor combinations that need to be considered, so for convenience we define 
$X_a:=q^{bc}\partial_{a}q_{bc}$ and $Y_a:=q^{bc}\partial_{c}q_{ab}$.
We evaluate the individual terms in \appref{sec:extras}.
When we substitute the results \eqref{eq:dtheta_components} into \eqref{eq:dist-eqn-sol}, we once again find a series of unique tensor combinations with their own coefficient which vanishes independently.  Most of these conditions are the same as those found in \eqref{eq:d2theta} so we won't bother duplicating them again here.  However, we do find the following new conditions,
\begin{subequations}
\begin{align}
    X^a \partial^b \theta_{ab} : 0 & =
    \partdif{ L }{ R } 
    - 4 \left( \partial_q \beta + 2 \beta \partial_q \right) \partdif{ L }{ w },
\label{eq:dtheta_1} \\
\begin{split}
    q^{ab} X^c \partial_c \theta_{ab} : 0 & =
    \frac{-1}{2} \partdif{ L }{ R }
    + \frac{ v }{ 3 } \left( 4 \partial_q - 1 \right)
    \partdif{^2 L }{ v \partial R }
    + \partdif{ \beta }{ v } \left( 1 - 2 \partial_q \right) \partdif{ L }{ v }
    + \left( \beta - 2 \partial_q \beta - 4 \beta \partial_q \right) \left( \partdif{^2 L }{ v^2 } - \frac{2}{3} \partdif{ L }{ w } \right),
\end{split}
\label{eq:dtheta_2} \\
\begin{split}
    v_\T^{ab} X^c \partial_c \theta_{ab} : 0 & =
    \frac{v}{3} \left( 4 \partial_q - 1 \right) \partdif{^2 L }{ w \partial R }
    + \partdif{ \beta }{ w } \left( 1 - 2 \partial_q \right) \partdif{ L }{ v }
    + \left( \beta - 2 \partial_q \beta - 4 \beta \partial_q \right) \partdif{^2 L }{ v \partial w },
\end{split}
\label{eq:dtheta_3}
\end{align}
\vspace{-5mm}
\begin{align}
    q^{ab} v_\T^{cd} X_d \partial_c \theta_{ab} : 0 & =
    \left( 1 - 2 \partial_q \right) \partdif{^2 L }{ v \partial R }
    - 4 \left( \partial_q \beta + 2 \beta \partial_q \right) \partdif{^2 L }{ v \partial w }
    - 4 \partdif{ \beta }{ v } \partial_q \partdif{ L }{ w },
\label{eq:dtheta_4} \\
    v_\T^{ab} v_\T^{cd} X_d \partial_c \theta_{ab} : 0 & =
    \left( 1 - 2 \partial_q \right) \partdif{^2 L }{ w \partial R } 
    - 4 \left( \partial_q \beta + 2 \beta \partial_q \right) \partdif{^2 L }{ w^2 }
    - 4 \partdif{ \beta }{ w } \partial_q \partdif{ L }{ w },
\label{eq:dtheta_5} \\
    q^{ab} v_\T^{cd} Y_d \partial_c \theta_{ab} : 0 & =
    2 \beta \partdif{^2 L }{ v \partial w }
    + \partdif{ \beta }{ v } \partdif{ L }{ w },
\label{eq:dtheta_6} \\
    v_\T^{ab} v_\T^{cd} Y_d \partial^c \theta_{ab} : 0 & =
    2 \beta \partdif{^2 L }{ w^2 }
    + \partdif{ \beta }{ w } \partdif{ L }{ w },
\label{eq:dtheta_7}
\end{align}
\vspace{-5mm}
\begin{align}
    \partial^a F \partial^b \theta_{ab} : 0 & = \left( \partdif{ \beta }{ F } + 2 \beta \partdif{}{ F } \right) \partdif{ L }{ w },
\label{eq:dtheta_8} \\
\begin{split}
    q^{ab} \partial^c F \partial_c \theta_{ab} : 0 & =
    \frac{2v}{3} \partdif{^3 L }{ F \partial v \partial R } 
    - \partdif{ \beta }{ v } \partdif{^2 L }{ F \partial v }
    - \left( \partdif{ \beta }{ F } + 2 \beta \partdif{}{F} \right) \left( \partdif{^2 L }{ v^2 } - \frac{2}{3} \partdif{ L }{ w } \right),
\end{split}
\label{eq:dtheta_9} \\
    v_\T^{ab} \partial^c F \partial_c \theta_{ab} : 0 & =
    \frac{2v}{3} \partdif{^3 L }{ F \partial w \partial R } 
    - \partdif{ \beta }{ w } \partdif{^2 L }{ F \partial v }
    - \left( \partdif{ \beta }{ F } + 2 \beta \partdif{}{F} \right) \partdif{^2 L }{ v \partial w },
\label{eq:dtheta_10}
\end{align}
\vspace{-5mm}
\begin{align}
    q^{ab} v_\T^{cd} \partial_d F \partial_c \theta_{ab} : 0 & =
    \half \partdif{^3 L }{ F \partial v \partial R } 
    + \partdif{ \beta }{ v } \partdif{^2 L }{ F \partial w }
    + \left( \partdif{ \beta }{ F } + 2 \beta \partdif{}{F} \right) \partdif{^2 L }{ v \partial w },
\label{eq:dtheta_11} \\
    v_\T^{ab} v_\T^{cd} \partial_d F \partial_c \theta_{ab} : 0 & =
    \half \partdif{^3 L }{ F \partial w \partial R } 
    + \partdif{ \beta }{ w } \partdif{^2 L }{ F \partial w }
    + \left( \partdif{ \beta }{ F } + 2 \beta \partdif{}{F} \right) \partdif{^2 L }{ w^2 },
\label{eq:dtheta_12}
\end{align}
    \label{eq:dtheta}%
\end{subequations}
where $F\in\{v,w,R\}$.

We have now acquired all conditions on the form of the Lagrangian for our choice of variables.  The next step is to try and consolidate them.


\section{Deriving the Lagrangian}
\label{sec:allact}

As shown in \secref{sec:intro_algebra}, the deformed Lagrangian must be calculated either perturbatively, as has been done in ref.~\cite{cuttell2014,cuttellthesis}, or completely generally.
Before we attempt the general calculation, we note the results found for the perturbative case which were derived in ref.~\cite{cuttellthesis} (though an incomplete form of the calculation was first done in ref.~\cite{cuttell2014}).  For a deformation function which depends quadratically on derivatives and an action which depends quartically on derivatives, we found that a deformed covariance was perturbatively maintained for the solutions,
\begin{subequations}
\begin{gather}
    \beta = \beta_\0 
    + \beta_{(R)} \left( 
    R + \frac{ \K }{ \beta_\0 } \right) 
    + \mathcal{O} \left( \partial q^3 \right).
        \label{eq:pert_sol_def}
\\
    L = L_\0 + \xi v \sqrt{q}
    + \frac{\omega}{2} \sqrt{ q \left| \beta_\0 \right| } \left\{ 
        R - \frac{ \K }{ \beta_\0 }
        - \frac{ \beta_{(R)} }{ 4 \beta_\0 } \left( R + \frac{ \K }{ \beta_\0 } \right)^2
    \right\}
    + \mathcal{O} \left( \partial q^5 \right),
        \label{eq:pert_sol_action}
\end{gather}%
    \label{eq:pert_sol}%
\end{subequations}
where $\K$ is what we call the standard extrinsic curvature contraction,
\begin{equation}
    \K = K^2 - K_{ab} K^{ab} = \quarter \left( v^2 - v_{ab} v^{ab} \right) = \frac{ v^2 }{ 6 } - \frac{ w }{ 4 }.
\end{equation}

We now turn to the calculation of the general deformed Lagrangian.
Take the equations \eqref{eq:d2theta} and \eqref{eq:dtheta}, which solve the distribution equation for the gravitational Lagrangian when we expand it in terms of the variables ${\{q,v,w,R\}}$, and see what can be deduced about the effective action when it is treated non-perturbatively.
Starting with the condition for ${\partial^aF\partial^b\theta_{ab}}$ where $F\in\{v,w,R\}$, \eqref{eq:dtheta_8}, this can be rewritten as
\begin{equation}
    0 = \beta \left( \partdif{ L }{ w } \right)^2 
    \partdif{}{ F } \log \left\{ \beta \left( \partdif{ L }{ w } \right)^2 \right\},
\end{equation}
which implies that
\begin{equation}
    \beta \left( \partdif{ L }{ w } \right)^2 = \lambda_1(q),
\end{equation}
and so we can solve up to a sign, ${\sigma_L=\pm1}$,
\begin{equation}
    \partdif{ L }{ w } = \sigma_L \sqrt{ \left| \frac{ \lambda_1 }{ \beta } \right| }. 
    \label{eq:dLdw}
\end{equation}
Then, from ${q^{ac}q^{bd}\partial_{cd}\theta_{ab}}$, \eqref{eq:d2theta_2}, we find
\begin{equation}
    \partdif{ L }{ R } = 4 \beta \partdif{ L }{ w } 
    = 4 \sigma_L \sigma_\beta \sqrt{ \left| \lambda_1 \beta \right| },
    \label{eq:dLdR}
\end{equation}
where ${\sigma_\beta:=\sgn{\beta(q,v,w,R)}}$.
If we then match the second derivative of the Lagrangian, 
${\partdif{^2L}{w\partial{R}}}$, 
using both equations, we find a nonlinear partial differential equation for the deformation function,
\begin{equation}
    0 = \partdif{ \beta }{ R } + 4 \beta \partdif{ \beta }{ w },
    \label{eq:def_pde}
\end{equation}
which is the same form as Burgers' equation for a fluid with vanishing viscosity \cite{smoller}.  However, before we attempt to interpret this, we will find further restrictions on the Lagrangian and deformation.

We now seek to find how the trace of the metric's normal derivative, $v$, appears.
Take the condition for ${v_\T^{ab}\partial^2\theta_{ab}}$, \eqref{eq:d2theta_4}
\begin{equation}
    0 = \frac{v}{3} \partdif{^2 L }{ R \partial w } - \beta \partdif{^2 L }{ v \partial w }
    = \frac{ \sigma_L }{ 2 } \sqrt{ \left| \frac{ \lambda_1 }{ \beta } \right| } \left( 
        \frac{ 4 v }{ 3 } \partdif{ \beta }{ w } + \partdif{ \beta }{ v } 
    \right)
\end{equation}
which we can solve to find that ${\beta=\beta\left(q,\bar{w},R\right)}$, where ${\bar{w}=w-2v^2/3}$. So in the deformation, the trace $v$ must always be paired with the traceless tensor squared $w$ like this.  We can see that this is related to the standard extrinsic curvature contraction by ${\bar{w}=-4\K}$.
To find how the trace appears in the Lagrangian, we look at the condition from ${q^{ab}\partial^2\theta_{ab}}$, \eqref{eq:d2theta_1},
\begin{equation}
    0 = \partdif{ L }{ R } - \frac{ 2 v }{ 3 } \partdif{^2 L }{ v \partial R }
    + 2 \beta \left( \partdif{^2 L }{ v^2 } - \frac{2}{3} \partdif{ L }{ w } \right).
\end{equation}
Substituting in our solutions so far, we can solve for the second derivative with respect to the trace,
\begin{equation}
    \partdif{^2 L }{ v^2 } = \frac{ - 4 \sigma_L }{ 3 } \sqrt{ \left| \frac{ \lambda_1 }{ \beta } \right| } \left(
        1 - \frac{ v }{ 2 } \partdif{ \beta }{ v }
    \right).
\end{equation}
We integrate over $v$ to find the first derivative,
\begin{equation}
    \partdif{ L }{ v } = \frac{ - 4 v \sigma_L}{ 3 } \sqrt{ \left| \frac{ \lambda_1 }{ \beta } \right| } + \xi_1 ( q, w, R )
    = \frac{ - 4 v }{ 3 } \partdif{ L }{ w } + \xi_1 ( q, w, R ),
        \label{eq:dLdv}
\end{equation}
where $\xi_1$ is a function arising as an integration constant.
To make sure that the solutions \eqref{eq:dLdw}, \eqref{eq:dLdR} and \eqref{eq:dLdv} match for the second derivatives 
${\partdif{^2L}{v\partial{R}}}$ and 
${\partdif{^2L}{v\partial{w}}}$, 
we find that ${\xi_1=\xi_1(q)}$.
Therefore, from this we can see that the Lagrangian should have the time derivatives only appear in the combined form $\bar{w}$ apart from a single linear term ${L\supset{}v\xi_1(q)}$.

Now we just have to determine what restrictions there are on how the metric determinant appears in the Lagrangian.
Firstly, we have the condition from ${X^a\partial^b\theta_{ab}}$, \eqref{eq:dtheta_1}, 
\begin{equation}
\begin{split}
    0 & = \partdif{ L }{ R } - 4 \left( \partial_q \beta + 2 \beta \partial_q \right) \partdif{ L }{ w },
\\ &
    = 4 \sigma_L \sigma_\beta \sqrt{ \left| \lambda_1 \beta \right| } \left( 1 - \frac{ \partial_q \lambda_1 }{ \lambda_1 } \right),
\quad
    \therefore \lambda_1 (q) = \lambda_2 q ,
\end{split}
\end{equation}
and secondly, from ${v_\T^{ab}X^c\partial_c\theta_{ab}}$, \eqref{eq:dtheta_3}, 
\begin{equation}
\begin{split}
    0 & = \frac{v}{3} \left( 4 \partial_q - 1 \right) \partdif{^2 L }{ w \partial R }
    + \partdif{ \beta }{ w } \left( 1 - 2 \partial_q \right) \partdif{ L }{ v }
    + \left( \beta - 2 \partial_q \beta - 4 \beta \partial_q \right) \partdif{^2 L }{ v \partial w },
\\ &
    = \partdif{ \beta }{ w } \left( \xi_1 - 2 \partial_q \xi_1 \right),
    \quad
    \therefore \xi_1 (q) = \xi_2 \sqrt{q}.
\end{split}
\end{equation}
Both these results show that our Lagrangian will indeed have the correct density weight when $\beta\to1$, that is $L\propto\sqrt{q}$.

All the remaining conditions from the distribution equation that have not been explicitly referenced are already solved by what we have found so far, so to make progress we must now attempt to consolidate our equations to find an explicit form for the Lagrangian.
If we integrate \eqref{eq:dLdw}, we find
\begin{equation}
    L = \sigma_L \sqrt{ \left| q \lambda_2 \right| } 
    \int_0^{\bar{w}} 
    \frac{ \mathrm{d} \bar{w}' }{ \sqrt{ \left| \beta ( q, \bar{w}', R ) \right| } }
    + f_1 ( q, v, R ),
        \label{eq:first_action}
\end{equation}
and then if we match the derivative of this with respect to $v$ with \eqref{eq:dLdv}, we find the $v$ dependence of the function which appeared as an integration constant,
\begin{equation}
    f_1 ( q , v , R ) = v \xi_2 \sqrt{q} + f_2 ( q , R ).
\end{equation}
If we then match the derivative of \eqref{eq:first_action} with respect to $R$ with \eqref{eq:dLdR}, we see that
\begin{equation}
    \partdif{ L }{ R }
    =
    4 \sigma_L \sigma_\beta \sqrt{ \left| q \lambda_2 \beta \right| } = \partdif{ f_2 }{ R }
    - \frac{\sigma_L}{2} \sqrt{ \left| q \lambda_2 \right| } \int_0^{\bar{w}} \frac{ \sigma_\beta \; \mathrm{d} \bar{w}' }{ \left| \beta( q , \bar{w}' , R ) \right|^{3/2} }
    \partdif{}{ R } \beta ( q, \bar{w}', R )
\end{equation}
and using \eqref{eq:def_pde} to change the derivative of $\beta$,
\begin{equation}
     4 \sigma_L \sigma_\beta \sqrt{ \left| q \lambda_2 \beta \right| }
     = \partdif{ f_2 }{ R }
    + 2 \sigma_L \sqrt{ \left| q \lambda_2 \right| } \int_0^{\bar{w}} \frac{ \mathrm{d} \bar{w}' }{ \sqrt{ \left| \beta ( q, \bar{w}', R ) \right| } }
    \partdif{}{\bar{w}'} \beta (q, \bar{w}', R ).
\end{equation}
from which we see we can change the integration variable,
\begin{equation}
    4 \sigma_L \sigma_\beta \sqrt{ \left| q \lambda_2 \beta \right| }
    = \partdif{ f_2 }{ R }
    + 2 \sigma_L \sqrt{ \left| q \lambda_2 \right| } 
    \int_{\beta(q,0,R)}^{\beta(q,\bar{w},R)} 
    \frac{ \mathrm{d} \beta' }{ \sqrt{|\beta'|} }.
\end{equation}
The upper integration limit cancels with the left hand side of the equality, and therefore
\begin{equation}
    \partdif{ f_2 }{ R } = 4 \sigma_L \sigma_0 \sqrt{ \left| q \lambda_2 \beta ( q, 0, R ) \right| }.
\end{equation}
where ${\sigma_0:=\sgn{\beta(q,0,R)}}$.
Then integrating this over $R$,
\begin{equation}
    f_2 ( q, R ) = 
    4 \sigma_L \sqrt{ \left| q \lambda_2 \right| } \int_0^R \sigma_0 \sqrt{ \left| \beta ( q, 0, R' ) \right| } \, \mathrm{d} R' + f_3 ( q ),
\end{equation}
which means that finally we have our solution for the general Lagrangian,
\begin{equation}
    L = \sigma_L \sqrt{ \left| q \lambda_2 \right| } \left(
        \int_0^{\bar{w}} \frac{ \mathrm{d} \bar{w}' }{ \sqrt{ \left| \beta ( q, \bar{w}', R ) \right| } }
        + 4 \int_0^R \sigma_0 \sqrt{ \left| \beta ( q, 0, R' ) \right| } \, \mathrm{d} R'
    \right)
    + v \xi_2 \sqrt{q}
    + f_3 (q).
\end{equation}
Now, we test this with a zeroth order deformation so we can match terms with our previous results. Using $\beta=\beta_\0(q)$,
\begin{equation}
    L = \sigma_L \sqrt{ \left| q \lambda_2 \right| } \left( 
        \frac{ \bar{w} }{ \sqrt{ \left| \beta_\0 \right| } }
        + 4 R \sgn{ \beta_\0 } \sqrt{ \left| \beta_\0 \right| }
    \right)
    + v \xi_2 \sqrt{q} + f_3 (q),
\end{equation}
comparing this to ref.~\cite{cuttell2018, cuttellthesis} and using $\bar{w}=-4\K$ leads to
\begin{equation}
    \sigma_L = \sigma_\beta,
\quad
    \sqrt{ \left| \lambda_2 \right| } = \frac{ \omega }{ 8 },
\quad
    f_3 = - \sqrt{q} \, U (q),
\end{equation}
and therefore, the full solution is given by,
\begin{equation}
\begin{split}
    L & = \frac{ \omega \sigma_\beta \sqrt{ q } }{ 2 } \left( 
        \int_0^R \sigma_0 \sqrt{ \left| \beta ( q, 0, R' ) \right| } \, \mathrm{d} R'
        - \int_0^{\K} \frac{ \mathrm{d} \K' }{ \sqrt{ \left| \beta ( q, \K', R ) \right| } }
    \right) 
    + \sqrt{q} \left( v \xi - U (q) \right),
        \label{eq:action}
\end{split}
\end{equation}
where we have relabeled ${\xi_2\to\xi}$, and the deformation function must satisfy the non-linear partial differential equation,
\begin{equation}
    \partdif{ \beta }{ R } = \beta \partdif{ \beta }{ \K }.
        \label{eq:def_pde2}
\end{equation}

By performing a Legendre transform, we can see that the Hamiltonian constraint associated with this Lagrangian \eqref{eq:action} is given by,
\begin{equation}
\begin{split}
    C & = \frac{ \omega \sigma_\beta \sqrt{q} }{ 2 }
    \left\{ 
        \int_0^{\K} \frac{ \mathrm{d} \K' }{ \sqrt{ \left| \beta ( q, \K', R ) \right| } }
        - \frac{ 2 \K }{ \sqrt{ \left| \beta ( q, \K, R ) \right| } }
        - \int_0^R \sigma_0 \sqrt{ \left| \beta ( q, 0, R' ) \right| } \, \mathrm{d} R'
    \right\}
    + \sqrt{q}\, U.
        \label{eq:constraint}
\end{split}
\end{equation}

\subsection{Solving for the deformation function}
\label{sec:allact_def}

The nonlinear partial differential equation for the deformation function \eqref{eq:def_pde2} is an unexpected result, and invites a comparison to a very different area of physics.
We can compare it to Burgers' equation for nonlinear diffusion, \cite{smoller},
\begin{equation}
    \partdif{ u }{ t } + u \partdif{ u }{ x } = \eta \partdif{^2 u }{ x^2 },
        \label{eq:burgers}
\end{equation}
(where $u$ is a density function), and see that our equation is very similar to the `inviscid' limit of vanishing viscosity $\eta\to0$.
This equation is not trivial to solve because it can develop discontinuities where the equation breaks down, termed `shock waves'.
Returning to our own equation \eqref{eq:def_pde2}, we analyse its characteristics. It implies that there are trajectories parameterised by $s$ given by
\begin{equation}
    \totdif{ q }{ s } = 0,
\quad
    \totdif{ R }{ s } = 1,
\quad
    \totdif{ \K }{ s } = - \beta \left( q, \K, R \right),
\end{equation}
along which $\beta$ is constant.  These trajectories have gradients given by,
\begin{equation}
    \totdif{ R }{ \K } = \frac{-1}{\beta\left(q,\K,R\right)}
\end{equation}
and because $\beta$ is constant along the trajectories, they are a straight line in the $(\K,R)$ plane.  We must have an `initial' condition in order to solve the equation, and because $R$ is here the analogue of $-t$ in \eqref{eq:burgers} we define the initial function when $R=0$, given by $\beta(q,\K,0)=:\alpha(q,\K)$.
Since there are trajectories along which $\beta$ is constant, we can use $\alpha$ to solve for $R(\K)$ along those curves, given an initial value $\K_0$,
\begin{equation}
    R = \frac{ \K_0 - \K }{ \alpha ( \K_0 ) }.
\end{equation}
We reorganise to get $\K_0=\K+R\alpha(\K_0)$, and then substitute into $\beta$.  This leads to the implicit relation,
\begin{equation}
    \beta ( q, \K, R ) = \alpha \left( q,
        \K + R \beta ( q, \K, R )
    \right).
\end{equation}
We invoke the implicit function theorem to calculate the derivatives of $\beta$,
\begin{equation}
    \partdif{ \beta }{ \K } = \frac{ \alpha' }{ 1 - R \alpha' },
\quad
    \partdif{ \beta }{ R } = \frac{ - \beta \alpha' }{ 1 - R \alpha' },
\end{equation}
which show that a discontinuity develops when $R\alpha'\to1$.  This is the point where the characteristic trajectories along which $\beta$ is constant converge to form a caustic. Beyond this point, $\beta$ seems to become a multi-valued function.

\begin{figure}[t]
	\begin{center}
	{\subfigure[$\beta$]{
		\label{fig:def_num_tanhbeta}
		\includegraphics[width = 0.3\textwidth]{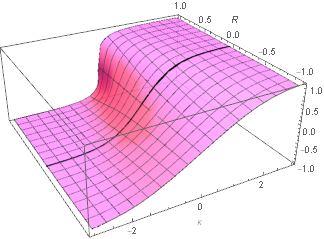}}}
	{\subfigure[$\partial\beta/\partial\K$]{
		\label{fig:def_num_tanhdbetadk}
		\includegraphics[width = 0.3\textwidth]{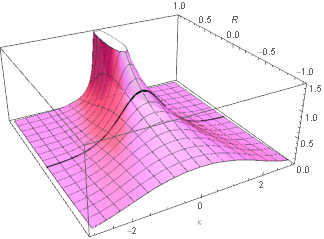}}}
	{\subfigure[$\partial\beta/\partial{R}$]{
		\label{fig:def_num_tanhdbetadr}
		\includegraphics[width = 0.3\textwidth]{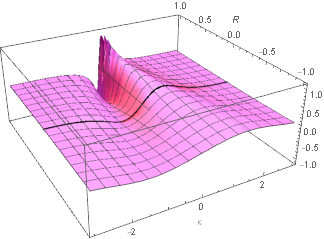}}}
	\end{center}
\caption[Numerically solved deformation with initial function \texorpdfstring{$\alpha=\tanh{\omega\K}$}{tanh(K)}]
{Numerically solved deformation function for initial function $\alpha=\tanh{(\omega\K)}$.  The numerical evolution breaks for $\omega{R}>1$ because a discontinuity has developed.  The initial function is indicated by the black line. The plots are in $\omega=1$ units.
}
	\label{fig:def_num_tanh}
\end{figure}

An analytic solution to $\beta$ only exists when $\alpha$ is linear in $\K$,
\begin{equation}
    \alpha = \alpha_\0 (q) + \alpha_{2} (q) \K,
\quad
    \beta = \frac{ \alpha_\0 (q) + \alpha_{2} (q) \K }{ 1 - \alpha_{2} (q) R },
        \label{eq:def_sol}
\end{equation}
which matches the equations for linear $\beta(\K,0)$ and the corresponding $\beta(0,R)$ found in ref.~\cite{cuttellconstraint}.
When ${|\alpha_{2}R|\ll1}$, we can expand $\beta$ into a series,
\begin{equation}
    \beta \simeq \alpha_\0 + \alpha_{2} \left( \K + \alpha_\0 R \right) \sum_{n=0}^{\infty} R^n \alpha_{2}^n,
        \label{eq:def_sol_expand}
\end{equation}
and by comparing this to the perturbative deformation found in ref.~\cite{cuttellthesis}, and written in equation \eqref{eq:pert_sol} we can see the correspondence $\alpha_\0=\beta_\0$ and 
$\alpha_{2}=\beta_{(R)}/\beta_\0$.
For non-linear initial functions, the deformation must be found numerically.  As a test, in \figref{fig:def_num_tanh}, we numerically solve for $\beta$ when $\alpha=\tanh{(\omega\K)}$.  We see that, as $R$ increases, the positive gradient in $\K$ intensifies to form a discontinuity, and softens as $R$ decreases.

\begin{figure}[t]
	\begin{center}
	{\subfigure[$\beta$]{
		\label{fig:def_num_cosbeta}
		\includegraphics[width = 0.333\textwidth]{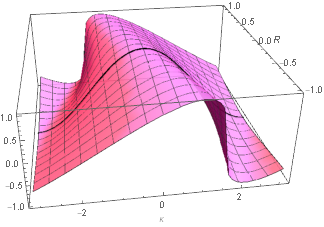}}}
	{\subfigure[$\partial\beta/\partial\K$]{
		\label{fig:def_num_cosdbetadk}
		\includegraphics[width = 0.333\textwidth]{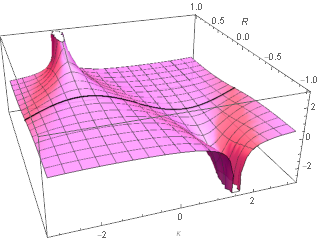}}}
\\
	{\subfigure[$\partial\beta/\partial{R}$]{
		\label{fig:def_num_cosdbetadr}
		\includegraphics[width = 0.333\textwidth]{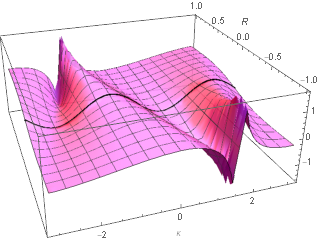}}}
	{\subfigure[$\beta$ cross sections]{
		\label{fig:def_num_cosbetacross}
		\includegraphics[width = 0.333\textwidth]{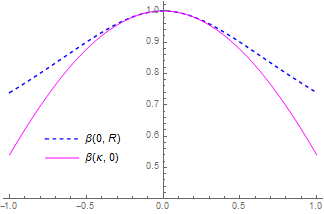}}}
	\end{center}
\caption[Numerically solved deformation with initial function \texorpdfstring{$\alpha=\cos{(\omega\K)}$}{cos(K)}]
{Numerically solved deformation function with an initial function $\alpha=\cos{(\omega\K)}$ and periodic boundary conditions.  The numerical evolution breaks for $|\omega{}R|>1$ because discontinuities have developed.  The initial function is indicated by the black line.  The plots are in $\omega=1$ units.
}
	\label{fig:def_num_cos}
\end{figure}

We have also numerically solved for the deformation when the initial function is given by $\alpha=\cos{(\omega\K)}$, shown in \figref{fig:def_num_cos}.
This function is somewhat motivated by loop quantum cosmology models with holonomy corrections \cite{Cailleteau2012a, Mielczarek:2012pf, Cailleteau2013}.  As with the $\tanh$ numerical solution in \figref{fig:def_num_tanh}, we see the positive gradient intensify and the negative gradient soften as $R$ becomes more positive.  We could not evolve the equations past the formation of the shock wave so from this we cannot determine for certain whether or not a periodicity emerges in $R$, but we can compare the cross sections for $\beta$ in \figref{fig:def_num_cosbetacross}.  
This cross section appears to match that found in ref.~\cite{cuttellconstraint} when we attempted to find the correspondence between $\beta(\K,0)$ and $\beta(\R)$ for non-linear functions. In that, $\R$ is a function of the canonical metric momentum and $R$.  It would seem that $\beta(0,R)$ should be a non-vanishing function even when $\beta(\K,0)$ does vanish for some values of $\K$.

When the inviscid Burgers' equation is being simulated in the context of fluid dynamics, a choice must be made on how to model the shock wave \cite{smoller}.  The direct continuation of the equation means that the density function $u$ becomes multi-valued, and the physical intepretation of it as a density breaks down.  The alternative is to then propagate the shock wave as a singular object, which requires a modification to the equations.  

Considering our case of the deformation function, allowing a shock wave to propagate does not seem to make sense.  It might require being able to interpret $\beta$ as a density function and the space of ${\left(\K,R\right)}$ to be interpreted as a medium. 
Whether or not the shock wave remains singular or becomes multi-valued, the most probable interpretation is that it represents a disconnection between different branches of curvature configurations.  That is, for a universe to transition from one side of the discontinuity to the other may require taking an indirect path through the phase space.
It may be that the behaviour in ${\left(\K,R\right)}$ is an embodiment of the curved momentum space hypothesis \cite{AmelinoCamelia:2011bm}.


\subsection{Linear deformation}
\label{sec:allact_linear}

If we take the analytic solution for the deformation function when its initial condition in linear \eqref{eq:def_sol}, we can substitute it into the general form for the gravitational Lagrangian \eqref{eq:action}.  If we assume we are in a region where $1-\alpha_{2}R>0$, we get the solution,
\begin{equation}
\begin{split}
    L &= \frac{ \omega \sqrt{q} }{ \alpha_{2} } \left\{
        \mathrm{sgn} \left( 1 + \frac{ \alpha_{2} \K }{ \alpha_\0 } \right) \sqrt{ \left| \alpha_\0  \right| }
        - \sqrt{ \left| \alpha_\0 + \alpha_{2} \K \right| }
        \sqrt{ 1 - \alpha_{2} R }
    \right\}
    + \sqrt{q} \left( v \xi - U \right),
        \label{eq:action_linear}
\end{split}
\end{equation}
and expanding in series for small derivatives of the metric when we are in a region where $|\alpha_\0|\gg|\alpha_{2}\K|$,
\begin{equation}
    L = \frac{ \omega }{ 2 } \sqrt{ q \left| \alpha_\0 \right| } \left(
        R - \frac{ \K }{ \alpha_\0 }
        - \frac{ \alpha_{2} }{ 4 } \left( R + \frac{ \K }{ \alpha_\0 } \right)^2
        + \mathcal{O} \left( \partial q^5 \right)
    \right)
    + \sqrt{q} \left( v \xi - U \right),
        \label{eq:action_linear_pert}
\end{equation}
which matches exactly the fourth order perturbative Lagrangian \eqref{eq:pert_sol} when ${\alpha_\0=\beta_\0}$ and ${\alpha_{2}=\beta_{(R)}/\beta_\0}$.

The Hamiltonian constraint associated with the non-perturbative effective action can be found by using \eqref{eq:constraint}.  Substituting in the Lagrangian for a linear deformation \eqref{eq:action_linear}, we can solve for $\K$ when the constraint vanishes (as long as we specify that it must be finite in the limit $\alpha_{2}\to0$),
\begin{equation}
    \K = \left\{
        \frac{ 2 }{ \omega } \sgn{ \alpha_\0 } \sqrt{ \left| \alpha_\0 \right| } U \left(
            1 - \frac{ \alpha_{2} U }{ 2 \omega \sqrt{ \left| \alpha_\0 \right| } }
        \right)
        - \alpha_\0 R
    \right\}
    \left( 1 - \frac{ \alpha_{2} U }{ \omega \sqrt{ \left| \alpha_\0 \right| } }
    \right)^{-2}.
\end{equation}
If we restrict to the FLRW metric, where ${q_{ab}=a^2\Sigma_{ab}}$, ${R=6ka^{-2}}$, ${\K=6\H^2}$, and ${U=\rho(a)}$, as described in ref.~\cite{cuttellconstraint}, we find the modified Friedmann equation,
\begin{equation}
    \H^2 = 
    \left\{
        \frac{ \sgn{ \alpha_\0 } \sqrt{ \left| \alpha_\0 \right| } }{ 3 \omega } \rho \left(
            1 - \frac{ \alpha_{2} \rho }{ 2 \omega \sqrt{ \left| \alpha_\0 \right| } }
        \right)
        - \frac{ \alpha_\0 k }{ a^2 }
    \right\}
        \left( 1 - \frac{ \alpha_{2} \rho }{ \omega \sqrt{ \left| \alpha_\0 \right| } }
    \right)^{-2}.
    \label{eq:friedmann_linear}
\end{equation}
There is a correction term similar to that found for the fourth order perturbative Lagrangian \cite{cuttell2014} which suggests there could be a bounce when $\rho\to2\omega\sqrt{\left|\alpha_\0 \right|}/\alpha_{2}$.  However, there is also an additional factor which causes $\H$ to diverge when $\rho\to\omega\sqrt{\left|\alpha_\0\right|}/\alpha_{2}$, which is before that potential bounce. 

\begin{sloppypar}
This is directly comparable to the modified friedmann equation found for the deformation function ${\beta(\R)=\beta_\0\left(1+\beta_2\R\right)^{-1}}$, investigated in ref.~\cite{cuttellconstraint}, with 
${\alpha_\0=\beta_\0}$ and
${\alpha_{2}=\omega\beta_2/2}$.  
As is found here, those results suggested a sudden singularity where $\H$ diverges when $a$ and $\rho$ remain finite.
Note that this is for the deformation function with a linear dependence on $\K$ which, unlike the cosine deformation, is not motivated by loop quantum cosmology.  It does, however, demonstrate the difference that higher order corrections can have on dynamics.
\end{sloppypar}
%


\section{Conclusions}
\label{sec:conclusions}

We derived the general deformed effective gravitational action from the deformed constraint algebra.  The way the deformation function is differently affected by extrinsic and intrinsic curvature (i.e. by time and space derivatives) was found to be similar to a differential equation which usually appears in fluid mechanics.  Discontinuities in the deformation function seem to be inevitable, but the interpretation of what they mean is not clear.  The discontinuities might be avoided if there were natural restrictions on the sign of the deformation's coefficients or the curvature.  This effect may be linked to the curved phase space hypothesis.

One of the original motivations of this study was to provide insight into the problem of incorporating spatial inhomogeneities into models of loop quantum cosmology which deform space-time covariance.  From our results, we can see that it is indeed possible to determine the dependence of the deformation on spatial derivatives from its dependence on extrinsic curvature.  However, the lack of analytical solutions, and numerical solutions which tend towards discontinuities, means that determining general behaviour is difficult.

That being said, there are important caveats to this work which must be kept in mind.  The fact that we used metric variables rather than the preferred connection or loop variables might limit the applicability of our results when comparing to the motivating theory.  Moreover, the deformation of the constraint algebra is only predicted for real values of $\BI$.
We also only considered combinations of derivatives that were a maximum of two orders, when higher order tensor combinations of derivatives and higher order derivatives are likely to appear in true quantum corrections.


\section*{Acknowledgements}
\label{sec:ackowledgements}

We would like to thank Martin Bojowald for invaluable help during this study.
MS is supported in part by the Science and Technology Facility Council (STFC), UK under the research grant ST/P000258/1.
The work of RC is supported by an STFC studentship.


\appendix

\section{Extra calculations }
\label{sec:extras}

For convenience, we use the definitions,
\begin{equation}
    X_a = q^{bc} \partial_a q_{bc},
\quad
    Y_a = q^{bc} \partial_c q_{ba} = \partial^b q_{ab},
\quad
    Z_a = v_\T^{bc} \partial_a q_{bc},
\quad
    W_a = v_\T^{bc} \partial_c q_{ba}.
        \label{eq:tensor_combinations}
\end{equation}
Evaluating each term in the $\partial_{cd}\theta_{ab}$ bracket of \eqref{eq:dist-eqn-sol}, by substituting in the variables 
\begin{equation}
    q := \det{q_{ab}},
\quad
    v := q^{ab} v_{ab},
\quad
    w := v^\T_{ab} v_\T^{ab} = v_{ab}v^{ab} - \third v^2,
\quad
    R := q^{bc} R^a_{\;\;bac}
\end{equation}
and using the equations derived for decomposing $R$ in ref.~\cite{cuttellconstraint},
\begin{subequations}
\begin{align}
    \partdif{ L }{ q_{ab,cd} }
    & = \left( Q^{abcd} - q^{ab} q^{cd} \right) \partdif{ L }{ R },
\\ 
    v_{ef} \partdif{^2 L }{ q_{ef,cd} \partial v_{ab} } 
    & =
    \left( v_\T^{cd} - \frac{2}{3} v q^{cd} \right) \left( 
        q^{ab} \partdif{^2 L }{ v \partial R } 
        + 2 v_\T^{ab} \partdif{^2 L }{ w \partial R } 
    \right),
\\
\begin{split}
    \partdif{^2 L }{ v_{ab} \partial v_{cd} }
    & =
    q^{ab} q^{cd} \left( \partdif{^2 L }{ v^2 } - \frac{2}{3} \partdif{ L }{ w } \right)
    + 2 Q^{abcd} \partdif{ L }{ w }
    + 2 \left( q^{ab} v_\T^{cd} + v_\T^{ab} q^{cd} \right) \partdif{^2 L }{ v \partial w }
    + 4 v_\T^{ab} v_\T^{cd} \partdif{^2 L }{ w^2 }.
\end{split}
\end{align}%
    \label{eq:d2theta_components}%
\end{subequations}

Evaluating each term in the $\partial_{c}\theta_{ab}$ bracket of \eqref{eq:dist-eqn-sol},
\begin{subequations}
\begin{equation}
\begin{split}
    \partdif{ L }{ q_{ab,c} } & = 
    \partdif{ L }{ R } \left(
        \frac{3}{2} Q^{abde} \partial^c q_{de} 
        - q^{c(d} q^{e)(a} \partial^{b)} q_{de}
        + q^{ab} Y^c
        - \half q^{ab} X^c
        - 2 q^{c(a} Y^{b)}
        + q^{c(a} X^{b)}
    \right),
\end{split}
\end{equation}
\begin{equation}
\begin{split}
    v_{ef} \partdif{^2 L }{ q_{ef,c} \partial v_{ab} } &=
    \left( 
        \frac{3}{2} Z^c - W^c - 2 v_\T^{cd} Y_d + v_\T^{cd} X_d + \frac{v}{3} X^c
    \right)
    \left( 
        q^{ab} \partdif{^2 L }{ v \partial R } 
        + 2 v_\T^{ab} \partdif{^2 L }{ w \partial R } 
    \right),
\end{split}
\end{equation}
\begin{equation}
\begin{gathered}
    v_{ef} \partial_d \left( \partdif{^2 L }{ q_{ef,cd} \partial v_{ab} } \right)  =
    \left( Z^c - W^c + \frac{v}{3} X^c + \frac{v}{3} Y^c - v_\T^{cd} Y_d \right) \left( q^{ab} \partdif{^2 L }{ v \partial R } + 2 v_\T^{ab} \partdif{^2 L }{ w \partial R } \right)
\\
    + \left( v_\T^{cd} - \frac{2v}{3} q^{cd} \right) \left\{
        \left( q^{ab} \partial_d - Q^{abef} \partial_d q_{ef} \right) \partdif{^2 L }{ v \partial R }
        + 2 \left( v_\T^{ab} \partial_d + Q^{abef} \partial_d v^\T_{ef} - 2 v_\T^{e(a} q^{b)f} \partial_d q_{ef} \right) \partdif{^2 L }{ w \partial R }
    \right\} ,
\end{gathered}
\end{equation}
\begin{equation}
\begin{split}
    \Gamma^c_{de} \partdif{^2 L }{ v_{ab} \partial v_{de} } & =
    \left( 2 q^{cd} q^{e(a} \partial^{b)} q_{de} - Q^{abde} \partial^c q_{de} \right) \partdif{ L }{ w }
    + \left( 2 W^c - Z^c \right) \left( q^{ab} \partdif{^2 L }{ v \partial w } + 2 v_\T^{ab} \partdif{^2 L }{ w^2 } \right)
\\ &
    + \left( Y^c - \half X^c \right) \left\{ q^{ab} \left( \partdif{^2 L }{ v^2 } - \frac{2}{3} \partdif{ L }{ w } \right) + 2 v_\T^{ab} \partdif{^2 L }{ v \partial w } \right\},
\end{split}
\end{equation}
\begin{equation}
\begin{split}
    \Gamma^c_{de} \partdif{ \beta }{ v_{ab} } \partdif{ L }{ v_{cd} } & =
    \left( q^{ab} \partdif{ \beta }{ v } + 2 v_\T^{ab} \partdif{ \beta }{ w } \right) \left\{
        \left( Y^c - \half X^c \right) \partdif{ L }{ v }
        + \left( 2 W^c - Z^c \right) \partdif{ L }{ w }
    \right\},
\end{split}
\end{equation}
\begin{equation}
\begin{split}
    \partial_d \beta \partdif{^2 L }{ v_{ab} \partial v_{cd} } & =
    \partial^c \beta \left\{ 
        q^{ab} \left( \partdif{^2 L }{ v^2 } - \frac{2}{3} \partdif{ L }{ w } \right) + 2 v_\T^{ab} \partdif{^2 L }{ v \partial w }
    \right\}
    + 2 q^{c(a} \partial^{b)} \beta \partdif{ L }{ w }
    + 2 v_\T^{cd} \partial_d \beta \left( q^{ab} \partdif{^2 L }{ v \partial w } + 2 v_\T^{ab} \partdif{^2 L }{ w^2 } \right)
\end{split}
\end{equation}
\begin{equation}
\begin{gathered}
    \partial_d \left( \partdif{^2 L }{ v_{ab} \partial v_{cd} } \right) =
    \left( q^{ab} \partial^c - q^{ab} Y^c - Q^{abef} \partial^c q_{ef} \right) \left( \partdif{^2 L }{ v^2 } \! - \! \frac{2}{3} \partdif{ L }{ w } \right)
    \! + \! 2 \left( q^{c(a} \partial^{b)} \! - \! q^{c(a} Y^{b)} \! - \! q^{c(e} q^{f)(a} \partial^{b)} q_{ef} \right) \partdif{ L }{ w }
\\ 
    + 2 \left\{ q^{ab} \left( v_\T^{cd} \partial_d \! - \! v_\T^{cd} Y_d \! - \! W^c \! + \! q^{cd} \partial^e v^\T_{de} \right)
    + v_\T^{ab} \partial^c - v_\T^{ab} Y^c
    + Q^{abef} \left( \partial^c v^\T_{ef} \! - \! v_\T^{cd} \partial_d q_{ef} \right) \! - \! 2 v_\T^{e(a} q^{b)f} \partial^c q_{ef}
    \right\} \partdif{^2 L }{ v \partial w }
\\
    + 4 \left\{ v_\T^{ab} \left( v_\T^{cd} \partial_d - W^c - v_\T^{cd} Y_d + q^{cd} \partial^e v^\T_{de} \right)
    + Q^{abef} v_\T^{cd} \partial_d v^\T_{ef} - 2 v_\T^{e(a} q^{b)f} v_\T^{cd} \partial_d q_{ef}
    \right\} \partdif{^2 L }{ w^2 },
\end{gathered}
\end{equation}
\begin{equation}
\begin{split}
    \partdif{ \beta }{ v_{ab} } \partial_d \left( \partdif{ L }{ v_{cd} } \right) & = 
    \left( q^{ab} \partdif{ \beta }{ v } + 2 v_\T^{ab} \partdif{ \beta }{ w } \right) 
    \left\{ 
        \left( \partial^c - Y^c \right) \partdif{ L }{ v }
        + 2 \left( v_\T^{cd} \partial_d + q^{cd} \partial^e v^\T_{de} - v_\T^{cd} Y_d - W^c \right) \partdif{ L }{ w }
    \right\}.
\end{split}
\end{equation}%
    \label{eq:dtheta_components}%
\end{subequations}
%


\bibliographystyle{./utphys}  
\bibliography{bibliography}  


\end{document}